\documentclass[12pt]{article}
\usepackage{amsfonts}
\usepackage{amsmath}
\begin{document}
\newtheorem{theorem}{Theorem}
\newtheorem{lemma}{Lemma}
\newtheorem{example}{Example}
\newtheorem{definition}{Definition}
\newtheorem{corollary}{Corollary}
\newtheorem{proposition}{Proposition}
\newcommand{\ord}{\mathop{\mathrm{ord}}\nolimits}
\newcommand{\es}{\mathop{\mathrm{ess\,sup}}\nolimits}
\newcommand{\Ord}{\mathop{\mathrm{Ord}}\nolimits}
\newcommand{\Id}{\mathop{\mathrm{Id}}\nolimits}
\newcommand{\Tr}{\mathop{\mathrm{Tr}}\nolimits}
\newcommand{\supp}{\mathop{\mathrm{supp}}\nolimits}
\newcommand{\id}{\mathop{\mathrm{Id}}\nolimits}
\newcommand{\Lip}{\mathop{\mathrm{Lip}}\nolimits}
\newcommand{\Qp}{\mathop{\mathbb Q_p}\nolimits}
\newcommand{\Zp}{\mathop{\mathbb Z_p}\nolimits}
\newcommand{\sgn}{\mathop{\mathrm{sgn}}\nolimits}
\newcommand{\esssup}{\mathop{\mathrm{ess\,sup}}\nolimits}
\newcommand{\rank}{\mathop{\mathrm{rank}}\nolimits}
\newcommand{\dist}{\mathop{\mathrm{dist}}\nolimits}
\newcommand{\pv}{\mathop{\mathrm{p.v.}}\nolimits}
\righthyphenmin=2

\title{$p$-Adic quantum calculus \\ and ideals of compact operators}  
\author{Evgeny I. Zelenov}
\date{ May 10, 2022} 
\maketitle

\begin{abstract}
The paper proposes a construction of a quantum differentiation operator defined on the spaces of complex-valued functions of  $p$-adic argument and taking values in the algebra of bounded operators on a Hilbert space. The properties of this operator are investigated. In particular, it is proved that the differentiation operator maps $p$-adic Besov spaces into Schatten-von Neumann ideals in the algebra of compact  operators.
\end{abstract}

\section{Introduction}
First of all, we will give a description of what will be called the $p$-adic quantum calculus. Consider the algebra $A=L^\infty (\mathbb Z_p,\mathbb C)$ of almost everywhere bounded measurable on $\mathbb Z_p$ functions taking value in the field of complex numbers. Let's define an irreducible representation of this algebra by multiplication operators in the Hilbert space $L^2=L^2(\mathbb Z_p)$, $A\ni f\mapsto M_f\in \mathcal B(L^2)$,
$$
M_f\colon L^2\ni h\mapsto fh\in L^2.
$$
On the space $L^2$, the Hilbert operator $S$ is defined (we will define it later), which is a self-adjoint operator and $S^2=\Id$.  For each function $f\in A$, we define a quantum differentiation operator
$$
df = [M_f\,,\,S].
$$
The correspondence $f\mapsto df$ constructed in this way between functions on $\mathbb Z_p$ and the operators $df$ on $L^2$  we will call $p$-adic quantum calculus.

The group $\mathbb Z_p$ of $p$-adic integers  is the inverse limit of finite cyclic groups $\mathbb Z/p^n\mathbb Z, \,n\in \mathbb Z_+$:
$$
\longleftarrow\mathbb Z/p^n\mathbb Z\longleftarrow\mathbb Z/p^{n+1}\mathbb Z\longleftarrow
$$
In this sense, the proposed construction is a combinatorial analogue of the quantum calculus suggested by A.\,Connes\cite{Connes}.

The purpose of this article is to study this correspondence. In particular, the following statements will be proved.
\begin{itemize}
\item
$df$ is a finite rank operator if and only if the function $f$ is locally constant;
\item
$df$ is a compact operator if and only if the function $f$ belongs to the space $VMO(\mathbb Z_p)$;
\item
$df$ belongs to the Schatten-von Neumann ideal $\mathfrak S^q$ if and only if $f$ belongs to the $p$-adic Besov space $B^{1/q}_{q,q}$.
\end{itemize}
Note that the last two statements are completely analogous  to the corresponding statements for algebras of complex-valued functions on the circle $\mathbb T\subset\mathbb C$.

The Connes quantum calculus (another name is the Connes quantization) has many interesting applications (see, for example, \cite{Sergeev1, Sergeev2}).

There is an important feature in the $p$-adic case considered in this paper. Namely, in all cases interesting for the application, there is no nontrivial "classical" differentiation (by differentiation on the algebra $A$ we mean a linear map $A\to A$ satisfying the Leibniz rule). The reason for this is that there are no nontrivial differentiations on the algebra of locally constant functions, and this algebra is a dense subalgebra of the algebra of observables in most interesting cases (\cite{VV,VVZ}) (for example, locally constant functions form a dense subset in the algebra of continuous on $\mathbb Z_p$ functions). At the same time, "quantum" differentiation, that is, differentiation with values in the algebra of bounded linear operators on a Hilbert space, is not trivial.

One of the possible applications of this paper is the construction of differential calculus on spaces of functions defined on totally disconnected spaces (\cite{BM, BP, KM}).

\section{p-Adic numbers, functions, spaces}
In this section, the necessary notations are introduced and the functional spaces in which we will work  are defined. Everywhere further we consider the case of $p\neq 2$.

Any nonzero $p$-adic number can be represented in the following form (canonical decomposition):
$$
\Qp\ni x=\sum_{k=-n}^{+\infty}x_kp^k, \,n\in\mathbb Z_+,\,x_k\in\{0,1,\dots, p-1\}.
$$
The finite part of the canonical decomposition, consisting of the sum of terms with negative powers of $p$, is the fractional part of $x$ and is denoted as $\{x\}_p$, respectively, the remaining part of the decomposition (with non-negative powers of $p$) is the integer part of $x$. Note that the fractional part is always a rational number, $p$-adic numbers with a trivial fractional part are $p$-adic integers $\mathbb Z_p$.
$$
\underbrace{p^{-n}x_{-n} + p^{-n+1}x_{-n+1}+\cdots +p^{-1}x_{-1}}_{\{x\}_p} + \underbrace{x_0+px_1+\cdots+p^{k}x_{k}+\cdots}_{[x]_p}
$$
Let's define the order of a nonzero $p$-adic number $x$ by the following rule:
$
\ord(x) = \min\{k\colon x_k\neq 0\}.
$
Order has the following property:
$
\ord(x+y)\geq\min\{\ord(x),\ord(y)\}.
$
$p$-adic norm is determined by the formula:
$
|x|_p = p^{-\ord(x)}.
$
This norm is non-Archimedean, that is, a strong triangle inequality is valid:
$
|x+y|_p\leq\max\{|x|_p\,,\, |y|_p\}.
$
By $\chi$ we denote the additive character of the field $\mathbb Q_p$ of the following form:
$\chi (x) = \exp\left(2\pi i\{x\}_p\right).$

For a nonzero $p$-adic number $x\in\mathbb Q_p^*$, the symbol $x_{\ord(x)}\in\{0,1,\dots p-1\}$ denotes the coefficient at the first nonzero term of the canonical decomposition of $x$. Let's define the sign $\sgn(x)$ of a nonzero $p$-adic number as the Legendre symbol of $x_{\ord(x)}$:
$\sgn(x) = \left(\frac{x_{\ord(x)}}{p}\right)$. The function $\sgn(x)$ takes values in the set $\{-1,1\}$ and is the multiplicative character of the field $\mathbb Q_p$:
$\mathbb Q^*_p\to\{-1,1\},\,\sgn(xy) = \sgn(x)\sgn(y)$.
A disk centered at zero of radius $p^{-n}$ is denoted by $B_n$:
$B_n = \{x\in\Zp\,\colon\,\ord(x)\geq n\,\equiv\,|x|_p\leq p^{-n}\}$.
In the following, we will consider functions defined on the ring $\mathbb Z_p$ of  $p$-adic integers  (as a set, $\mathbb Z_p$ is a unit disk, $\mathbb Z_p=B_0$) and taking values in the field $\mathbb C$ of complex numbers.

A function $f$ is locally constant of order $n$ if it is invariant with respect to shifts: $f(x+u) = f(x),\, x\in\Zp,\, u\in B_n$. The set of such functions form a finite-dimensional vector space $LC_n$ over complex numbers of dimension $p^n$. The basis vectors of this space are indicator functions of disks of radius $p^{-n}$ in $\mathbb Z_p$. Since the topological space $\mathbb Z_p$ is totally disconnected, locally constant functions are continuous. The following embeddings are valid:
$m\leq n, \,LC_m\subseteq LC_n$. The set of all locally constant functions will be denoted by $LC$. This space has a natural structure of the direct limit of finite-dimensional subspaces: $LC = \varinjlim LC_n$. The Haar measure on $\mathbb Z_p$ will be denoted by $\mu$. Lebesgue spaces of measurable functions with respect to this measure are defined in a standard way:
$ L^q\,\equiv\,\{f\colon \Zp\to \mathbb C,\, \|f\|_q = \left(\int_{\Zp}|f|^qd\mu\right)^{1/q}<\infty\},\,
\|f\|_\infty = \esssup_{\Zp}|f|$.

There is a chain of embeddings:
$$LC\subset L^\infty\subset\,\cdots\,\subset L^q\subset\,\cdots\,\subset L^1,$$
$LC$ is dense in $L^q$ for all $1\leq q\leq\infty$.

\section{Hilbert Transform}
In this section, we define and consider the basic properties of the $p$-adic Hilbert transform (or operator). A large number of papers have been devoted to the study of the properties of this type of operators, here are just some, see \cite{F1, T, ChT}.

The Hilbert transform $S$ of the function $f$ is defined by the following formula (the integral is understood in the sense of principal value):
$$
((Sf)(x) = \frac{1}{\Gamma}\,\pv\int_{\Zp}\frac{\sin(x-y)}{|x-y|_p}f(y)dy,\,\,
\Gamma = \begin{cases}\sqrt p, & p=1(\mod 4)\\ i\sqrt p, & p=3(\mod 4)\end{cases}
$$
We use the notation $S$ for the Hilbert operator, since in the future it plays the role of a symmetry operator. The Hilbert operator is  bounded  from $L^q$ to $L^q$ for all $1<q<\infty$. For $f\in L^1$, the estimate is valid:
$$\mu\left\{x\in\Zp\,\colon\,|(Sf)(x)|>\lambda\right\}\leq C\frac{\|f\|_1}{\lambda}.$$
\section{Fourier transform and Hilbert transform}
The following section presents some well-known facts from harmonic analysis over the field of $p$-adic numbers \cite{T}.

Any character of the group $\mathbb Z_p$ of $p$-adic integers has the form $\chi_\alpha(x) = \chi(\alpha x), \,x\in\mathbb Z_p,\,\alpha\in\mathbb Q_p$. Thus, the Pontryagin dual of the group $\mathbb Z_p$ is $\mathbb Q_p/\mathbb Z_p$:
$$
\Hat{\mathbb Z}_p = \left\{\chi_\alpha(x) = \chi(\alpha x),\,\,\alpha\in\Qp/\Zp\right\}.
$$
In the last formula in the expression $\chi(\alpha x)$ an arbitrary representative from the corresponding adjacency class is taken as $\alpha$.
Since the function $\chi(x)$ is identically equal to 1 on $\mathbb Z_p$, then the formula  is correct. The group $\Hat{\mathbb Z}_p$ is also known as the Prufer group $\mathbb Z(p^\infty)$, it is a quasi-cyclic group and can be represented as a subgroup in $\mathbb T$ consisting of $p^n$-th roots of  unity $n\in\mathbb N$.

The set of functions
$\left\{\chi_\alpha(x),\,\alpha\in\Hat{\mathbb Z}_p\right\}$ forms an orthonormal basis in $L^2$. The Fourier transform of the function $\phi\in L^2$ is defined in the standard way:
$$
\phi\in L^2,\,F[\phi](\alpha) = \langle\phi,\chi_\alpha\rangle = \hat\phi_\alpha = \int_{\Zp}\phi(x)\chi(-\alpha x)dx.
$$
Note that the function $\chi_\alpha$ is a locally constant of order $n$, where $n$ is determined by the norm $\alpha$: $\chi_\alpha\in LC_n, \,\,|\alpha|_p = p^n$. The norm $|\cdot|_p$ can be correctly defined on $\Qp/\Zp$, assuming it is equal to zero on a zero adjacency class and equal to the norm of any representative from a non-zero  adjacency class (due to the non-archimedean norm, such a definition does not depend on the choice of a representative). Moreover, the family of functions $\left\{\chi_\alpha(x),\,\alpha\in\Hat{\mathbb Z}_p,\,|\alpha|_p\leq p^n\right\}$ forms an orthonormal basis in the space $LC_n$. It also follows that the function $\phi$ is locally constant if and only if it is represented as a finite Fourier series:
$\phi\in LC\iff \Hat\phi_\alpha=0$ for almost all $\alpha\in\Hat{\Zp}$.

In this sense, locally constant functions are analogs of trigonometric polynomials. However, there is an important difference - the ratio of two locally constant functions (provided that the function in the denominator does not take zero values) is again a locally constant function.

Hilbert transform has the following properties..
\begin{itemize}
\item
The Hilbert transform maps constant  functions into a function identically equal to zero:
$\left(S\mathbb I_{\Zp}\right)(x) = 0$.
\item
The characters $\chi_\alpha,\,\alpha\in\Hat{\mathbb Z}_p$ are eigenfunctions of the Hilbert operator with eigenvalues $\sgn(\alpha)$:
$\left(S\chi_\alpha\right)(x) = \sgn(\alpha)\chi_\alpha(x),\,\alpha\in\Hat{\mathbb Z}_p,\,x\in\Zp$. It should be noted here that the function $\sgn$ can be correctly defined on the group $\Hat{\mathbb Z}_p$ as follows: $\sgn(\alpha)$ is set equal to zero on the unit element in $\Hat{\mathbb Z}_p$ and equal to the value on an arbitrary adjacency class representative otherwise. Obviously, this value does not depend on the choice of a representative.
\item
As can be easily seen from the previous property of the Hilbert operator, this operator is self-adjoint, and its square is equal to the identity operator. That is, the Hilbert operator is a symmetry operator:
$S^*=S, \,\, S^2=1$.
\item
The space $L^2_0$ (the space of functions from $L^2$ orthogonal to constants) can be decomposed into a direct orthogonal sum of subspaces $W^+$ and $W^-$: $L^2=W^+\bigoplus W^-$, the space $W^+$ is linear span of the basis vectors $\chi_\alpha$, for which the condition $\sgn(\alpha)=1$ is met, the space $W^-$ is linear span of the basis vectors $\chi_\alpha$, for which the condition $\sgn(\alpha)=-1$ is met. By $P^+$ and $P^-$ we denote orthogonal projectors into subspaces $W^+$ and $W^-$, respectively. Then the representation is valid for the Hilbert operator
$S=P^+-P^-$.
\end{itemize}

\section{Differentiation operator}
In this section, we will define the quantum differentiation operator in the $p$-adic case (hereinafter - the differentiation operator), and give its simplest properties.
\begin{definition}
Let $f\in L^\infty$ and $M_f$ be the multiplication operator in $L^2$, $\left(M_f\phi\right)(x) = f(x)\phi(x)$. The operator in $L^2$ of the following form
$$
df = \left[M_f\,,\,S\right]
$$
will be called the value of the differentiation operator of the function $f$.
\end{definition}
The square brackets in the last formula denote the commutator. Thus, the differentiation operator is defined on the algebra $L^\infty$ of functions on $\Zp$ and takes a value in the algebra $\mathcal B = \mathcal B(L^2)$ of bounded linear operators on the space $L^2$. It is obvious that the differentiation operator satisfies the Leibniz rule: $d(fg)=(df)g+f(dg)$. We will also call the value of the operator $d$ on the function $f$ the derivative of the function $f$.

It immediately follows from the definition that the derivative of a constant function is a trivial operator on $\mathcal B$ (maps any function to the identically zero function). Let's give a less trivial example.
\begin{example}
\label{1}
Let $f(x)= \chi_a(x)$. Let us find the derivative of this function. Since $df$ is an operator in $L^2$, we calculate the value of this operator on the basis vector $\chi_\alpha$. The answer is as follows:
$$\left(d\chi_a\right)\chi_\alpha = \big(\sgn(\alpha)-\sgn(\alpha+a)\big)\chi_{\alpha+a}.$$
\end{example}
The following equations are valid:
\begin{multline*}
\left(d\chi_a\right)\chi_\alpha = \left[M_{\chi_a}\,,\,S\right]\chi_\alpha = \chi_aS\chi_\alpha - S(\chi_a\chi_{\alpha}) = \\ =\chi_a\sgn(\alpha)\chi_\alpha - \sgn(\alpha+a)\chi_{\alpha+a} = \big(\sgn(\alpha)-\sgn(\alpha+a)\big)\chi_{\alpha+a}.
\end{multline*}
The following theorem is valid.
\begin{theorem}
Let $f\in L^\infty$. The derivative $df$ is an operator of finite rank if and only if the function $f$ is locally constant.
\end{theorem}
Let $f\in LC$. Then, as noted above, the function $f$ is represented as a finite linear combination of characters $\chi_a$, that is, $f(x) = \sum_{a\in\Omega}\hat f_a\chi_a(x)$, while $a$ belongs to a finite subset of $\Omega$ in $\Hat{\mathbb Z}_p$. Therefore, the inequality $\rank(df) \leq\sum_{a\in\Omega}\rank(d\chi_a)$ is valid. Consider the operator $T=d\chi_a(d\chi_a)^*$. The rank of the operator $T$ coincides with the rank of the operator $d\chi_a$. As follows from the Example \ref{1}, the operator $T$ acts on the basis vectors $\chi_\alpha$ according to the following formula:
\begin{equation}
\label{T}
T\chi_\alpha= \left(\sgn(\alpha+a)-\sgn(\alpha)\right)^2\chi_\alpha.
\end{equation}
Thus, the characters $\chi_\alpha$ are the eigenvalues of the operator $T$, the corresponding eigenvalues are $$\lambda_\alpha = \left(\sgn(\alpha+a)-\sgn(\alpha)\right)^2.$$ Let $a\neq 0$. Then it is easy to see that $\lambda_0 = \lambda_{-a} = 1$. In addition, as follows from the definition of the function $\sgn$, $\sgn(\alpha+a) = \sgn(\alpha)$ for all $\alpha$ such that the inequality $|\alpha|_p>|a|_p$ is satisfied. Thus, the image of the operator $T$ is a subspace of the linear span of the basis vectors $\chi_\alpha$, for which the condition $|\alpha|_p\leq|a|_p$ is satisfied. The dimension of this subspace does not exceed $|a|_p$. Hence, $\rank(d\chi_a)\leq |a|_p$. Hence the sufficiency of the condition of the theorem is proved.

It is possible to obtain an exact formula for the rank of the operator $d\chi_a$. Namely:
\begin{equation}
\label{rank1}
\rank(d\chi_a) = \frac{|a|_p+3}{2}.
\end{equation}
Indeed, in a finite field of order $p^n$, the number of nonzero elements that are squares coincides with the number of elements that are not squares and is equal to $1/2(p^n-1)$. Therefore, for $\alpha\neq 0$ and $\alpha\neq-a$, we have $1/2(|a|_p-1)$ of nonzero eigenvalues of $\lambda_\alpha$, and we need to add two more, namely, $\lambda_0$ and $\lambda_{-a}$.
If the function $f\in L^\infty$ is not locally constant, then its Fourier series expansion contains nonzero terms $\hat f_a\chi_a$ with an index $a$ of arbitrarily large norm. Taking into account the formula (\ref{rank1}), the operator $df$ cannot have a finite rank in this case.
\begin{corollary}
\label{cont}
If the function $f$ is continuous, $f\in C(\Zp)$, then its derivative $df$ is a compact operator.
\end{corollary}
Indeed, the mapping $f\mapsto M_f$ is continuous in a uniform topology on $C(\Zp)$ and norm topologies on $\mathcal B(L^2)$. Mapping $A\mapsto [A,S],\,A\in\mathcal B(L^2)$ is a differentiation on the algebra of bounded operators $\mathcal B(L^2)$ and is therefore continuous in the topology of the norm (\cite{Sakai}). Thus, the differentiation operator $f\mapsto df$ is continuous in a uniform topology on $C(\Zp)$ and norm topologies on $\mathcal B(L^2)$. Since locally constant functions are dense in $C(\Zp)$, the image of the algebra of continuous functions lies in the closure of the space of operators of finite rank in the topology of the norm in $\mathcal B(L^2)$, that is, in the algebra $\mathcal K(L^2)$ of compact operators on $L^2$. Corollary \ref{cont} gives a sufficient condition for the compactness of the derivative, however, this condition is not necessary.
\section{Hilbert-Schmidt operators}
Let's define the $p$-adic Sobolev space $H^{1/2}$. This space is an analogue of the Sobolev space of semi-differentiable functions. To do this, on the space $LC$ of locally constant functions, we define the $\|\cdot\|_{1/2}$ semi-norm as follows. Let $f\in LC$ be represented as its (finite) Fourier series:
$f(x)=\sum_{a\in\Hat{\mathbb Z}_p}\Hat f_a\chi_a(x)$. Then, by definition,
\begin{equation}
\label{sobolev}
\|f\|_{1/2} = \left(\sum_{a\in\Hat{\mathbb Z}_p}|a|_p|\Hat f_a|^2\right)^{1/2}.
\end{equation}
The kernel of this semi-norm is a set of constant functions.
The space $H^{1/2}$ is defined as the closure of the space $LC$ with respect to the seminorm (\ref{sobolev}).
\begin{lemma}
Let $f(x)=\sum_{a\in\Hat{\mathbb Z}_p}\Hat f(a)\chi_a(x)$. The following formula is valid:
$$
\Tr\left(df\right)^* \left(df\right) = 2\sum_{a\in\Hat{\mathbb Z}_p}|a|_p|\Hat f_a|^2.
$$
\end{lemma}
Let's use the formula (\ref{T}) to calculate the diagonal matrix element of the operator $\left(df\right)^*\left(df\right)$:

\begin{multline}
\label{tr}
\langle\left(df\right)^*\left(df\right)\chi_{\alpha},\chi_\alpha\rangle = \\ = \left\langle\sum_{a,b\in\Hat{\mathbb Z}_p}\hat {f}_a\Bar{\hat {f}}_b\left(\sgn(\alpha)-\sgn(a+\alpha)\right)\left(\sgn(\alpha)-\sgn(b+\alpha)\right)\chi_{\alpha+a-b},\chi_\alpha\right\rangle = \\ = \sum_{a\in\Hat{\mathbb Z}_p}|\hat {f}_a|^2\left(\sgn(\alpha)-\sgn(a+\alpha)\right)^2.
\end{multline}
As noted above, $\lambda_\alpha = \left(\sgn(\alpha)-\sgn(a+\alpha)\right)^2$ takes exactly two values equal to one (for $\alpha = 0$ and $\alpha = -a$) and the value 4 exactly for $1/2(|a|_p-1)$ of different values of $\alpha$, for the rest $\alpha$ the eigenvalues of $\lambda_\alpha$ are equal to zero. Therefore, taking into account the relations (\ref{tr}),
the equalities are valid:
\begin{multline*}
\Tr\left(df\right)^* \left(df\right) = \sum_{\alpha\in\Hat{\mathbb Z}_p}\sum_{a\in\Hat{\mathbb Z}_p}|\hat {f}_a|^2\left(\sgn(\alpha)-\sgn(a+\alpha)\right)^2 = \\=\sum_{a\in\Hat{\mathbb Z}_p}|\hat {f}_a|^2\left(2+4\frac{|a|_p-1}{2}\right) = 2\sum_{a\in\Hat{\mathbb Z}_p}|a|_p|\hat {f}_a|^2.
\end{multline*}
Thus, we have proved the following theorem.
\begin{theorem}
The derivative $df$ is a Hilbert-Schmidt operator, $df\in\mathfrak S^2$, if and only if the function $f$ belongs to the space $H^{1/2}$.
\end{theorem}

\section{Spaces $BMO(\Zp)$ and $VMO(\Zp)$}
Initially, we defined the differentiation operator for the functions $f\in L^\infty$. In fact, this operator can naturally be extended to a wider space. We will give definitions of the corresponding spaces. Let the function $f$ be integrable on $\Zp$, $f\in\L^1$ and $B$ denote an arbitrary disk in $\Zp$. Using $f_B$, we denote the average value of the function $f$ on the disk $B$:
$$f_B = \frac{1}{\mu(B)}\int_Bf(x)dx.$$
Let's calculate the average deviation of the function $f$ from its average value on the disk $B$, and consider the maximum of such average deviations for all disks of radius not exceeding $p^{-n}, \,n\in\mathbb Z_+$:
$$M_n = \max_{\{B\colon\mu(B)\leq p^{-n}\}}\frac{1}{\mu(B)}\int_{B}|f(x)-f_B|dx.$$
We will say that the function $f$ belongs to the space $BMO = BMO(\Zp)$ (bound mean oscillation), if the sequence $M_n$ is bounded, $\|f\|_{BMO} = \sup_nM_n$ sets a semi-norm on the BMO space, the kernel of this semi-norm consists of constant functions. A typical example of a function from the BMO space is the function $\log|x|_p$. A subspace of the BMO space consisting of functions for which the condition is met
$\lim_{n\to\infty}M_n=0$ denote $VMO =VMO(\Zp)$ (vanishing mean oscillation). The BMO space contains the space $L^\infty$ as its proper subspace and is the natural space of the definition of the differentiation operator.
\begin{theorem}
\label{BMO}
The derivative $df$ of the function $f$ is a bounded operator on $L^2$, $df\in\mathcal B(L^2)$, if and only if $f$ belongs to the space $BMO$. The derivative $df$ is a compact operator, $df\in\mathcal K(L^2)$, if and only if the function $f$ belongs to the VMO space.
\end{theorem}
These statements are completely analogous to the corresponding statement for quantum differentiation in the complex case and are proved in a similar way. Therefore, we will present only a sketch of the proof. It is known (\cite{T, DK}) that the Hilbert operator is a continuous operator from BMO to BMO and a continuous operator from $L^\infty$ to BMO. In addition, the following characterization of the BMO space is valid in terms of the action of the Hilbert operator on the space $L^\infty$. Namely, any function $f\in BMO$ is represented as $f=g+Sh$, where the functions $g$ and $h$ belong to the space $L^\infty$, while the inequalities $\|f\|_{BMO}\geq C\|g\|_\infty, \,\|f\|_{BMO}\geq C\|h\|_\infty$  are valid for some constant $C$. This result was proved for functions defined on $\mathbb R^N$ (\cite{FS}), for functions on $\Zp$ the proof is similar. Using this representation for a function from the BMO space, to continue the differentiation operator from $L^\infty$ to the BMO space, it is enough to do this on  $SL^\infty$, and this follows from the continuity of the Hilbert operator from $L^\infty$ to $BMO$.
Similarly, the second part of the theorem can be proved. Instead of the result of Fefferman and Stein (\cite{FS}), we will use the following result of Sarason (\cite{S}), more precisely,   the $p$-adic analogue of this result. A function $f\in BMO$ belongs to the VMO space if and only if it can be represented as $f =g+Sh$, where the functions $g$ and $h$ are continuous on $\Zp$. Taking into account the last statement, the continuity of the Hilbert operator and Corollary 1 of the Theorem \ref{1}, we obtain the second statement of the theorem.

\section{Besov spaces $B_{q,r}^{s}$}
A considerable amount of work has been devoted to the Besov spaces of functions on the field  $\Qp$ of $p$-adic numbers (and more generally to the spaces of functions on profinite groups and martingales), for example \cite{OW,F,GT}.

Let $q,r,s$ be real numbers satisfying the inequalities $s>0,\,\, 1\leq q,r\leq\infty$. The subspace of functions from $L^1$ for which the seminorm $\|\cdot\|_{B_{q,r}^s}$ defined below is finite, we will call the Besov space $B_{q,r}^{s}$:
\begin{multline*}
B_{q,r}^{s} = \Big\{f\in L^1\,\colon\, \|f\|_{B_{q,r}^{s}} = \\
\left(\int_{\Zp}\left(\frac{\|f(x-y)-f(x)\|_q}{|y|_p^s}\right)^r \frac{dy}{|y|_p}\right)^{1/r} <\infty\Big\}.
\end{multline*}
An equivalent definition of the Besov space is usefull (\cite{OW}). We introduce the notation $\Delta_n(x)$ for the normalized indicator function of the disk $B_n$:
$\Delta_n(x)=\frac{1}{\mu\left(B_n\right)}\mathbb I_{B_n}(x)$. The mapping $L^1\ni f\mapsto f*\Delta_n\in LC_n$ defines the projection of the space $L^1$ into the space $LC_n$ ($*$ denotes convolution).
\begin{proposition}
Let the conditions be met: $f\in L^1,\,\,s>0,\,1\leq q,r\leq\infty$. The seminorm $\|f\|_{B_{q,r}^{s}}$ is equivalent to the following seminorm:
$$\|f\|_{B_{q,r}^{s}}\,\asymp\,
\left(\sum_{n=0}^{\infty}\left(p^{ns}\|f-f*\Delta_n\|_q\right)^r\right)^{1/r}.$$
\end{proposition}
Below, we will need the case $r=q, \,s=1/q$.
An obvious consequence follows from the last statement.
\begin{corollary} The function $f$ belongs to the Besov space $B_{q,q}^{1/q}$, $f\in B_{q,q}^{1/q}$, if and only if the condition is satisfied
$ p^{n/q}\|f-f*\Delta_n\|_q\in \ell^q$.
\end{corollary}
The connection of Besov and BMO spaces on Vilenkin groups was studied in \cite{OW}. In particular, a stronger statement follows from the results of this work. Namely:
\begin{proposition}
\label{est}
The function $f$ belongs to the Besov space $B_{q,q}^{1/q}$, $f\in B_{q,q}^{1/q}$, if and only if the following condition is satisfied
$ p^{n/q}\|f-f*\Delta_n\|_{BMO}\in \ell^q$.
\end{proposition}
The following theorem is valid.
\begin{theorem} The derivative $df$ belongs to the Schatten-von Neumann ideal $\mathfrak S^q$, $df\in\mathfrak S^q$, if and only if the function $f$ belongs to the Besov space $B_{q,q}^{1/q}$, $f\in B_{q,q}^{1/q}$.
\end{theorem}
Recall the definition of the ideal $\mathfrak S^q$. Let
$K\in\mathcal K(L^2)$ is a compact operator. The singular number $s_n(K)$ of this operator is defined as the distance in the operator norm to the space of operators of rank no higher than $n$:
$$s_n(K) = \inf_R\{\|K-R\|\,\colon\,\rank R \leq n\}.$$
By definition, $K\in\mathfrak S^q$ if the condition $\{s_n(K)\}\in\ell^q$ is satisfied. The last condition is equivalent to the following condition:$\{p^{n/q}s_{p^n}(K)\}\in\ell^q$. Since $f*\Delta_n\in LC_n$, the derivative $d(f*\Delta_n)$ is an operator of rank no higher than $p^n$ (Theorem \ref{1}). Therefore, the inequality $s_{p^n}(df)\leq\|d(f-f*\Delta_n)\|$  is valid. In fact, a two-way estimate of $s_{p^n}(df) \asymp\|d(f-f*\Delta_n)\|$ is valid, since the function $f*\Delta_n$ gives the best approximation in the topology of the VMO space of the function $f$ by a locally constant function from $LC_n$, and the differentiation operator is a continuous mapping from VMO to the algebra of compact operators. Taking into account  Theorem \ref{BMO} , the chain of relations is valid:
$$s_{p^n}(df)\asymp\|d(f-f*\Delta_n)\|\,\asymp\|f-f*\Delta_n\|_{BMO}.$$
Taking into account the Proposition \ref{est}, the statement of the theorem follows from the last chain of inequalities.

\end{document}